\documentclass[aps,pra,reprint,onecolumn,11pt,nofootinbib,showkeys]{revtex4-2}

\usepackage[T1]{fontenc}
\usepackage{newtxtext,newtxmath}
\usepackage{amsmath}
\usepackage{graphicx}

\usepackage{booktabs}
\usepackage{placeins}
\usepackage{microtype}
\usepackage{xcolor}
\usepackage[
colorlinks=true,
linkcolor=blue,
citecolor=blue,
urlcolor=blue,
bookmarksnumbered=true
]{hyperref}

\newtheorem{theorem}{Theorem}
\newtheorem{lemma}{Lemma}

\newenvironment{proof}{\par\noindent\textit{Proof.}\ }{\hfill$\square$\par}
\newcommand{\Tr}{\operatorname{Tr}}
\newcommand{\Id}{\mathbb{I}}
\newcommand{\HS}{\mathrm{HS}}

\newcommand{\Itilde}{\widetilde I_{\alpha}}
\newcommand{\Jtilde}{\widetilde J_{\alpha}}
\newcommand{\LB}{\mathrm{LB}}
\newcommand{\SU}{\mathrm{SU}}

\begin{document}

\title{Generalized Wigner-Yanase Uncertainty Bounds for Multiple Quantum Channels}

\author{Jianming Zhou}
\affiliation{School of Mathematics and Physics, University of South China, Hengyang, Hunan 421001, China}

\author{Xiaoli Hu}
\email[Corresponding author: ]{xiaolihumath@jhun.edu.cn}
\affiliation{School of Artificial Intelligence, Jianghan University, Wuhan, Hubei 430056, China}

\author{Naihuan Jing}
\affiliation{Department of Mathematics, North Carolina State University, Raleigh, North Carolina 27695, USA}

\begin{abstract}
We derive uncertainty relations for quantum channels based on the generalized Wigner-Yanase skew information. 
For three channels, we obtain a product lower bound by optimizing over permutations of commutator and anticommutator pairings. 
For an arbitrary number of channels, we establish a lower bound on the sum of squared uncertainties and construct a hierarchy of bounds using the partial Cauchy-Schwarz inequality. 
These bounds are invariant under changes of Kraus representation for a fixed operator basis. 
The first two levels of the hierarchy coincide with the elementary arithmetic-geometric mean bound, while the intermediate levels generally depend on the ordering of the operator basis. 
Examples involving three and four qubit channels demonstrate the bounds and compare their performance with an earlier pairing inequality under fixed channel representations.
\end{abstract}

\maketitle


\section{Introduction}\label{sec:introduction}

Uncertainty relations constrain the measurement statistics of incompatible quantum observables and have applications in entanglement detection, quantum cryptography, and quantum metrology \cite{Coles2017,Conlon2026}. 
Following Heisenberg's uncertainty principle, Kennard and Robertson established variance bounds for noncommuting observables \cite{Heisenberg1927,Kennard1927,Robertson1929}.
Entropic uncertainty relations provide complementary bounds on measurement outcome distributions \cite{Maassen1988,Berta2010,Tomamichel2011}.
For mixed states, however, the variance includes contributions from both quantum uncertainty and classical mixing \cite{Luo2006}. 
The Wigner-Yanase skew information characterizes the noncommutativity between a quantum state and an observable, providing a measure of the associated quantum uncertainty \cite{Wigner1963,Luo2003}.
This approach has been extended through the Wigner-Yanase-Dyson family and metric-adjusted skew information \cite{Furuichi2009,Hansen2008}.

Skew information is also closely related to quantum coherence and its behavior under quantum operations. 
The resource theory of coherence provides a framework for quantifying coherence in quantum states, and its connections with skew information have been investigated \cite{Baumgratz2014,Streltsov2017,LuoSun2017a,LuoSun2017b}. 
For a quantum channel $\Phi(\rho)=\sum_i K_i\rho K_i^\dagger$, the sum of the skew information associated with its Kraus operators defines a channel uncertainty measure that is independent of the Kraus representation \cite{Nielsen2010,Luo2018}.
Related measures based on metric-adjusted skew information and other operator constructions have also been studied \cite{SunLiLuo2022,Xuan2023,RenLuoLi2024}.
This notion of channel uncertainty characterizes the channel relative to a given input state, rather than the statistics of measurement outcomes associated with a quantum process.

Sum uncertainty relations based on Wigner-Yanase skew information were initially developed for quantum observables and subsequently extended to quantum channels \cite{Chen2016,Fu2019}.
Related bounds have also been established using metric-adjusted skew information \cite{Cai2021,Ren2021}.
Further studies obtained refined uncertainty bounds for observables and channels and extended the channel uncertainty relations to an arbitrary number of channels \cite{ZhangGaoYan2021,ZhangWuFei2021}.
Further channel uncertainty bounds have been developed using Wigner-Yanase skew information \cite{Zhou2023,Hu2023,ZhangWuFei2023,ZhangFei2024}.
Alternative approaches have employed modified generalized Wigner-Yanase-Dyson skew information and modified generalized variance to quantify channel uncertainty \cite{XuWuFei2022}.
Sum and product uncertainty relations based on metric-adjusted skew information have also been investigated \cite{ZhangWuMaFei2023,XuZhangFei2024}, together with channel uncertainty relations involving $\rho$-absolute variance, symmetrized $\rho$-absolute variance, and modified Wigner-Yanase skew information \cite{XuZhouZhangFei2024,XuZhangFei2024b}.
More recently, a channel uncertainty measure involving the arithmetic mean of $\rho^\alpha$ and $\rho^{1-\alpha}$ was introduced, leading to product, sum, and refined uncertainty relations for two quantum channels \cite{Wu2025}.

In this work, we study uncertainty relations for multiple quantum channels on a common Hilbert space of finite dimension using the generalized Wigner-Yanase skew information.
For three channels, a product uncertainty bound is obtained by optimizing the pairings of commutator and anticommutator terms over all permutations.
We first establish a bound on the sum of squared uncertainties for two channels and then extend the construction to arbitrary $N$, yielding a hierarchy of lower bounds through the partial Cauchy-Schwarz inequality.
All these bounds are independent of Kraus representations; the intermediate levels, however, can depend on the choice and ordering of the orthonormal operator basis.
For three and four qubit channels, we examine the bounds numerically and compare them with the commutator bound and the pairing bound of Ref.~\cite{Wu2025}, using the same input states and channel parameters with fixed Kraus representations.


\section{Cross channel uncertainty relations}\label{sec:relations}

Let $\rho$ be a density operator on a $d$-dimensional Hilbert space, and let $0<\alpha<1$.
For a quantum channel $\Phi$ with Kraus operators $\{K_i\}$, define $\rho_\alpha$ as an arithmetic mean and the centered Kraus operators $\widetilde K_i$ as follows:
\begin{align}
	\rho_\alpha &= \frac{\rho^\alpha+\rho^{1-\alpha}}{2}, \quad
	\widetilde K_i = K_i-\Tr(\rho K_i)\Id.
	\label{eq:center}
\end{align}
The positive operator $\rho_\alpha$ is not necessarily normalized, and $\rho_{1/2}=\sqrt{\rho}$.
Following Ref.~\cite{Wu2025}, define
\begin{align}
	\Itilde(\rho,\Phi) &= \frac12\sum_i\|[\rho_\alpha,\widetilde K_i]\|_{\HS}^{2},\\
	\Jtilde(\rho,\Phi) &= \frac12\sum_i\|\{\rho_\alpha,\widetilde K_i\}\|_{\HS}^{2},
	\label{eq:IJ}\\
	Q_\alpha(\rho,\Phi) &= \sqrt{\Itilde(\rho,\Phi)\Jtilde(\rho,\Phi)}.
	\label{eq:Q}
\end{align}
Here $\|A\|_{\HS}=\sqrt{\Tr(A^\dagger A)}$ denotes the Hilbert-Schmidt norm, while $[A,B]=AB-BA$ and $\{A,B\}=AB+BA$ are the commutator and anticommutator, respectively.
At $\alpha=1/2$, Eq.~\eqref{eq:Q} reduces to the centered Wigner-Yanase channel uncertainty of Ref.~\cite{ZhangWuFei2023}.

For channels $\Phi_s$ with Kraus operators $\{K_i^{(s)}\}_i$, write
\begin{align}
X_i^{(s)}&=[\rho_\alpha,\widetilde K_i^{(s)}],\\ Y_i^{(s)}&=\{\rho_\alpha,\widetilde K_i^{(s)}\},
\label{eq:XY}\\
C_{st}&=\sum_{i,j}\left|\Tr\left([K_j^{(t)},(K_i^{(s)})^\dagger]\rho_\alpha^2\right)\right|^2 .
\label{eq:C}
\end{align}
By taking complex conjugates and interchanging $i$ and $j$, one obtains $C_{st}=C_{ts}\geqslant0$.

\begin{lemma}\label{lem:mixed}
For all $i,j,s,t$, the mixed inner product satisfies
\begin{align}
\langle X_i^{(s)},Y_j^{(t)}\rangle_{\HS}
=\Tr\left([K_j^{(t)},(K_i^{(s)})^\dagger]\rho_\alpha^2\right).
\label{eq:mixed}
\end{align}
\end{lemma}
\begin{proof}
The centering scalar vanishes from the commutator and its contribution to the mixed inner product is zero because $\Tr([\rho_\alpha,K]^\dagger \rho_\alpha)=0$.
By cyclicity of the trace,
\begin{align}
\langle X_i^{(s)},Y_j^{(t)}\rangle_{\HS}
&=\Tr\left([(K_i^{(s)})^\dagger,\rho_\alpha]\{\rho_\alpha,K_j^{(t)}\}\right)\notag\\
&=\Tr\left(K_j^{(t)}(K_i^{(s)})^\dagger \rho_\alpha^2-(K_i^{(s)})^\dagger K_j^{(t)}\rho_\alpha^2\right),
\end{align}
which is Eq.~\eqref{eq:mixed}.
\end{proof}

\begin{theorem}\label{thm:product}
For any three quantum channels on the same $d$ dimensional Hilbert space, we have
\begin{align}
\prod_{s=1}^{3}Q_\alpha(\rho,\Phi_s)
&\geqslant\frac18\left[\max_{\pi\in S_3}\prod_{s=1}^{3}C_{s,\pi(s)}\right]^{1/2}.
\label{eq:product}
\end{align}
\end{theorem}
\begin{proof}
For every permutation $\pi\in S_3$, the Cauchy-Schwarz inequality and Lem.~\ref{lem:mixed} give
\begin{align}
\left(\prod_{s=1}^{3}Q_\alpha(\rho,\Phi_s)\right)^2
&=\frac1{64}\prod_{s=1}^{3}\left(\sum_i\|X_i^{(s)}\|_{\HS}^2\right)
\left(\sum_j\|Y_j^{(s)}\|_{\HS}^2\right)\notag\\
&\geqslant\frac1{64}\prod_{s=1}^{3}\sum_{i,j}|\langle X_i^{(s)},Y_j^{(\pi(s))}\rangle_{\HS}|^2\notag\\
&=\frac1{64}\prod_{s=1}^{3}C_{s,\pi(s)} .
\end{align}
Taking the maximum proves Eq.~\eqref{eq:product}.
\end{proof}

Because $C_{st}=C_{ts}$, the permutation maximum contains precisely the five candidate expressions
\begin{align}
\max_{\pi\in S_3}\prod_sC_{s,\pi(s)}
=\max\{C_{11}C_{22}C_{33},C_{12}^{2}C_{33},C_{13}^{2}C_{22},C_{23}^{2}C_{11},C_{12}C_{13}C_{23}\} .
\label{eq:product-candidates}
\end{align}
At $\alpha=1/2$, the last candidate gives the cyclic commutator pairing in the Wigner-Yanase channel setting.
Depending on the channels, a noncyclic permutation may yield the largest bound.

\begin{theorem}\label{thm:sum}
For any two channels $\Phi_s$ and $\Phi_t$, we have
\begin{align}
Q_\alpha(\rho,\Phi_s)^2+Q_\alpha(\rho,\Phi_t)^2&\geqslant\frac12C_{st} .
\label{eq:two-sum}
\end{align}
Consequently, for $N\geqslant2$,
\begin{align}
\sum_{s=1}^{N}Q_\alpha(\rho,\Phi_s)^2
&\geqslant\frac{1}{2(N-1)}\sum_{1\leqslant s<t\leqslant N}C_{st} .
\label{eq:N-sum}
\end{align}
\end{theorem}
\begin{proof}
By the arithmetic-geometric mean inequality and Cauchy-Schwarz,
\begin{align}
Q_\alpha(\rho,\Phi_s)^2+Q_\alpha(\rho,\Phi_t)^2
&\geqslant2\sqrt{\Itilde(\rho,\Phi_s)\Jtilde(\rho,\Phi_t)\Itilde(\rho,\Phi_t)\Jtilde(\rho,\Phi_s)}\notag\\
&\geqslant\frac12\sqrt{C_{st}C_{ts}}=\frac12C_{st} .
\end{align}
Summation over all unordered pairs and division by $N-1$ gives Eq.~\eqref{eq:N-sum}.
\end{proof}

\section{Partial Cauchy-Schwarz bounds}\label{sec:hierarchy}

Fix an ordered orthonormal Hilbert-Schmidt basis $\{E_k\}_{k=1}^{d^2}$ and identify an operator $X$ with the coordinate vector $x_k=\Tr(E_k^\dagger X)$.
For $x,y\in\mathbb C^{d^2}$ and integers $0\leqslant p\leqslant d^2$, set
\begin{align}
F_p(x,y)&=\|x\|^2\|y\|^2-\sum_{1\leqslant k<l\leqslant p}|x_ky_l-x_ly_k|^2 ,
\label{eq:F}\\
T_{st}^{(p)}&=\frac14\sum_{i,j}F_p\left(x_i^{(s)},y_j^{(t)}\right),
\label{eq:T}
\end{align}
where $x_i^{(s)}$ and $y_j^{(t)}$ are the coordinates of the operators in Eq.~\eqref{eq:XY}.
The empty sum implies $F_0=F_1$.
The complex Lagrange identity gives
\begin{align}
F_0(x,y)=F_1(x,y)\geqslant F_2(x,y)\geqslant\cdots\geqslant F_{d^2}(x,y)=|\langle x,y\rangle|^2 .
\label{eq:Lagrange}
\end{align}

\begin{theorem}\label{thm:hierarchy}
For $N\geqslant2$, each integer $0\leqslant p\leqslant d^2$ gives
\begin{align}
\sum_{s=1}^{N}Q_\alpha(\rho,\Phi_s)^2
\geqslant\LB_N^{(p)},
\label{eq:hierarchy}
\end{align}
where
\begin{align}
 \LB_N^{(p)}=\frac{2}{N-1}\sum_{s<t}\sqrt{T_{st}^{(p)}T_{ts}^{(p)}} .
\label{eq:LB-definition}
\end{align}
These bounds satisfy
\begin{align}
\LB_N^{(0)}=\LB_N^{(1)}\geqslant\LB_N^{(2)}\geqslant\cdots\geqslant\LB_N^{(d^2)}
=\frac{1}{2(N-1)}\sum_{s<t}C_{st} .
\label{eq:hierarchy-order}
\end{align}
\end{theorem}
\begin{proof}
Equations~\eqref{eq:IJ}, \eqref{eq:mixed}, and \eqref{eq:Lagrange} imply
\begin{align}
T_{st}^{(0)}&=\Itilde(\rho,\Phi_s)\Jtilde(\rho,\Phi_t),\quad T_{st}^{(d^2)}=\frac14C_{st},\quad T_{st}^{(p)}\geqslant T_{st}^{(p+1)} .
\end{align}
Applying the arithmetic geometric mean inequality to $Q_\alpha(\rho,\Phi_s)^2+Q_\alpha(\rho,\Phi_t)^2$ gives
\begin{align}
Q_\alpha(\rho,\Phi_s)^2+Q_\alpha(\rho,\Phi_t)^2
&\geqslant2\sqrt{T_{st}^{(0)}T_{ts}^{(0)}}
\geqslant2\sqrt{T_{st}^{(p)}T_{ts}^{(p)}} .
\end{align}
Summing over pairs proves Eq.~\eqref{eq:hierarchy}, while monotonicity and $C_{st}=C_{ts}$ give Eq.~\eqref{eq:hierarchy-order}.
\end{proof}

Write $u_s=Q_\alpha(\rho,\Phi_s)$.
At $p=0,1$, the bound and its gap are explicitly
\begin{align}
\LB_N^{(0)}=\LB_N^{(1)}&=\frac{2}{N-1}\sum_{s<t}u_su_t ,
\label{eq:elementary}\\
\sum_su_s^2-\LB_N^{(0)}&=\frac1{N-1}\sum_{s<t}(u_s-u_t)^2 .
\label{eq:gap}
\end{align}
Thus these two levels are an equality case of the elementary arithmetic geometric mean inequality when all $u_s$ coincide.
These levels provide no additional estimate once the individual $u_s$ are known.
The intermediate bounds $2\leqslant p<d^2$ lie between this elementary bound and the cross channel commutator bound.
The values at $p=0,1,d^2$ are independent of the ordered operator basis, whereas intermediate values need not be.

\begin{theorem}\label{thm:kraus}
For a fixed state $\rho$, a parameter $\alpha$, and an ordered operator basis, the quantities $C_{st}$ and $T_{st}^{(p)}$ are unchanged by arbitrary changes of the Kraus representations of the channels.
Consequently, all bounds in Theorems~\ref{thm:product}, \ref{thm:sum}, and \ref{thm:hierarchy} are independent of the Kraus representations.
\end{theorem}

\begin{proof}
	Let $\{A_{\mu}^{(s)}\}_{\mu=1}^{r_s}$ be a minimal Kraus representation of $\Phi_s$, where $r_s$ is the rank of its Choi operator.
	Every Kraus representation $\{K_i^{(s)}\}_{i=1}^{m_s}$ of $\Phi_s$ can be written as
	\begin{align}
		K_i^{(s)}=\sum_{\mu=1}^{r_s}V_{i\mu}A_{\mu}^{(s)},\quad V^\dagger V=\Id,
	\end{align}
	where $V\in\mathbb C^{m_s\times r_s}$.
	The same holds independently for $\Phi_t$.
	Any two Kraus representations can alternatively be related by a unitary transformation after padding the shorter list with zero operators.
    We may therefore consider the isometric changes
\begin{align}
	K_a^{\prime(s)} &= \sum_i V_{ai}K_i^{(s)}, \quad
	K_b^{\prime(t)} = \sum_j W_{bj}K_j^{(t)},
\end{align}
   where $V$ and $W$ satisfy $V^\dagger V=W^\dagger W=\Id$.
   For the matrix $c$ with entries
	\begin{align}
		c_{ij}=\Tr\bigl([K_j^{(t)},(K_i^{(s)})^\dagger]\rho_\alpha^2\bigr),
	\end{align}
the transformation gives
	\begin{align}
		c'=\overline VcW^{\mathsf T},\quad
		\|c'\|_{\mathrm F}=\|c\|_{\mathrm F}.
	\end{align}
	Hence, $C_{st}$ in Eq.~\eqref{eq:C} is independent of the Kraus representations.
	
	Under the same transformations, the operator coordinates satisfy
	\begin{align}
		x_a^{\prime(s)}&=\sum_iV_{ai}x_i^{(s)},\quad
		y_b^{\prime(t)}=\sum_jW_{bj}y_j^{(t)}.
	\end{align}
	Since $V$ and $W$ are isometries, the sums of squared norms are invariant, and consequently so is
	\begin{align}
		\sum_{i,j}\|x_i^{(s)}\|^2\|y_j^{(t)}\|^2.
	\end{align}
	For each $k<l$, define the matrix $D^{kl}$ by
	\begin{align}
		D_{ij}^{kl}
		=(x_i^{(s)})_k(y_j^{(t)})_l
		-(x_i^{(s)})_l(y_j^{(t)})_k.
	\end{align}
	Under Kraus isometries,
	\begin{align}
		D^{\prime kl}=VD^{kl}W^{\mathsf T},\quad
		\|D^{\prime kl}\|_{\mathrm F}=\|D^{kl}\|_{\mathrm F}.
	\end{align}
	Therefore, $T_{st}^{(p)}$ in Eq.~\eqref{eq:T} is independent of the Kraus representations for every fixed $p$.
\end{proof}


\section{Examples}\label{sec:examples}

We evaluate the inequalities for the mixed qubit state $\rho(\theta)=\frac12(\Id+\boldsymbol r\cdot\boldsymbol\sigma)$, with $\boldsymbol r=\frac{\sqrt3}{2}(\cos\theta,\sin\theta,0)$.
The three channels are amplitude damping, bit flip, and phase damping, with Kraus operators
\begin{align}
\mathcal E_{\mathrm{AD}}(q):\;& A_1=|0\rangle\langle0|+\sqrt{1-q}\,|1\rangle\langle1|,\quad A_2=\sqrt q\,|0\rangle\langle1|,\notag\\
\mathcal E_{\mathrm{BF}}(q):\;& B_1=\sqrt q\,\Id,\quad B_2=\sqrt{1-q}\,\sigma_x,\notag\\
\mathcal E_{\mathrm{PD}}(q):\;& D_1=|0\rangle\langle0|+\sqrt{1-q}\,|1\rangle\langle1|,\quad D_2=\sqrt q\,|1\rangle\langle1| .
\label{eq:channels-qubit}
\end{align}
Here $0\leqslant q\leqslant1$.
For $N=4$, we add the depolarizing channel $\mathcal E_{\mathrm D}(q)$ with Kraus operators
\begin{align}
\bigl\{\sqrt{1-q}\,\Id,\,\sqrt{q/3}\,\sigma_x,\,\sqrt{q/3}\,\sigma_y,\,\sqrt{q/3}\,\sigma_z\bigr\} .
\label{eq:depolarizing}
\end{align}
For intermediate bounds we use the ordered orthonormal operator basis
$\bigl(|0\rangle\langle0|,\,|1\rangle\langle0|,\,|0\rangle\langle1|,\,|1\rangle\langle1|\bigr) .$
The Kraus operators retain the ordering in Eq.~\eqref{eq:channels-qubit}.

For a comparison at fixed Kraus representations with the pairing bound of Ref.~\cite{Wu2025}, define
\begin{align}
\Lambda_{st}&=\frac12\left(\sum_i|\langle X_i^{(s)},X_i^{(t)}\rangle_{\HS}|\right)
\left(\sum_j|\langle Y_j^{(s)},Y_j^{(t)}\rangle_{\HS}|\right) .
\label{eq:Lambda}
\end{align}
When needed, shorter Kraus lists are padded with zero operators.
This expression uses the commutator and anticommutator pairings of Theorem~2 in Ref.~\cite{Wu2025}.
The same inequality holds for all $0<\alpha<1$, since the Cauchy-Schwarz and arithmetic-geometric mean inequalities give
\begin{align}
\Lambda_{st}\leqslant2Q_\alpha(\rho,\Phi_s)Q_\alpha(\rho,\Phi_t)
\leqslant Q_\alpha(\rho,\Phi_s)^2+Q_\alpha(\rho,\Phi_t)^2 .
\label{eq:Lambda-bound}
\end{align}
Unlike $C_{st}$ and $T_{st}^{(p)}$, the quantity $\Lambda_{st}$ may depend on the Kraus lists and their ordering.
We denote the uniform pairing and hybrid bounds by
\begin{align}
B_{\Lambda}&=\frac1{N-1}\sum_{s<t}\Lambda_{st},\quad B_{\mathrm{hyb}}=\frac1{N-1}\sum_{s<t}\max\left\{\frac12C_{st},\Lambda_{st}\right\} .
\label{eq:comparison}
\end{align}
No general ordering between these pairing bounds and the intermediate levels $\LB_N^{(p)}$ is established here.

\begin{figure}[tbp]
	\centering
	
	\begin{minipage}[b]{0.42\linewidth}
		\centering
		\includegraphics[width=\linewidth]{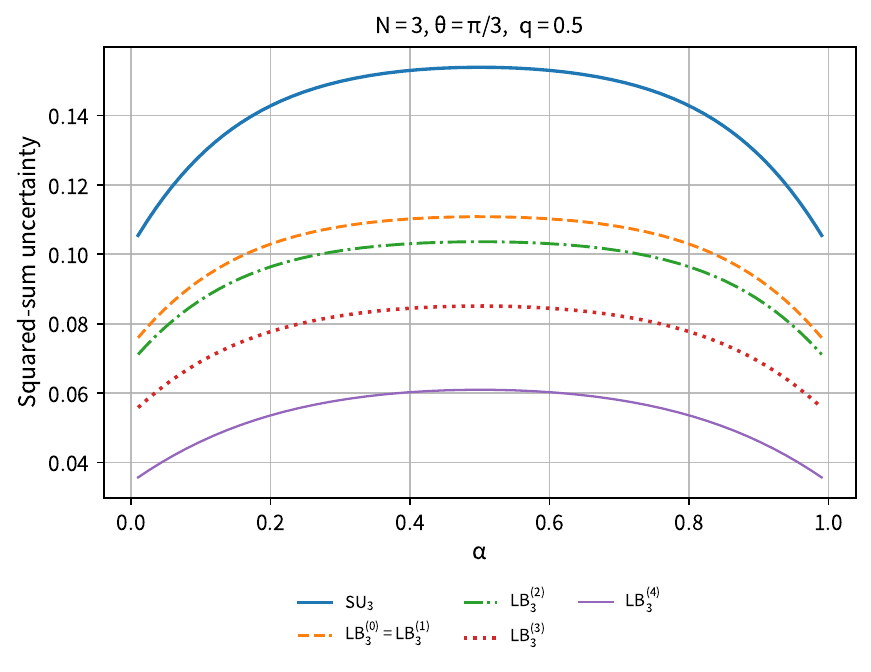}
		\par\smallskip
		\small (a) $N=3$, varying $\alpha$
	\end{minipage}
	\hfill
	\begin{minipage}[b]{0.42\linewidth}
		\centering
		\includegraphics[width=\linewidth]{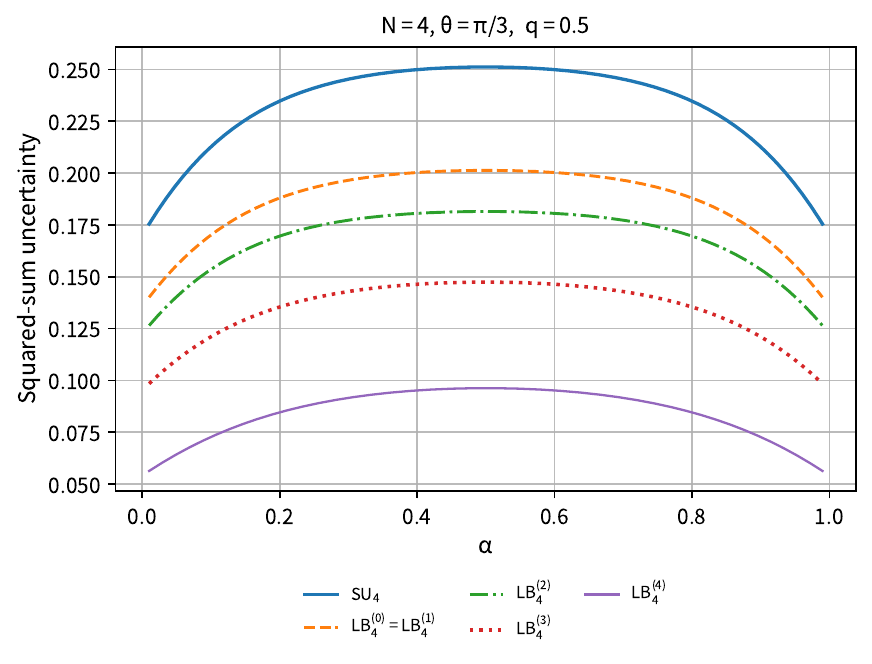}
		\par\smallskip
		\small (b) $N=4$, varying $\alpha$
	\end{minipage}
	
	\vspace{0.3em}
	
	\begin{minipage}[b]{0.42\linewidth}
		\centering
		\includegraphics[width=\linewidth]{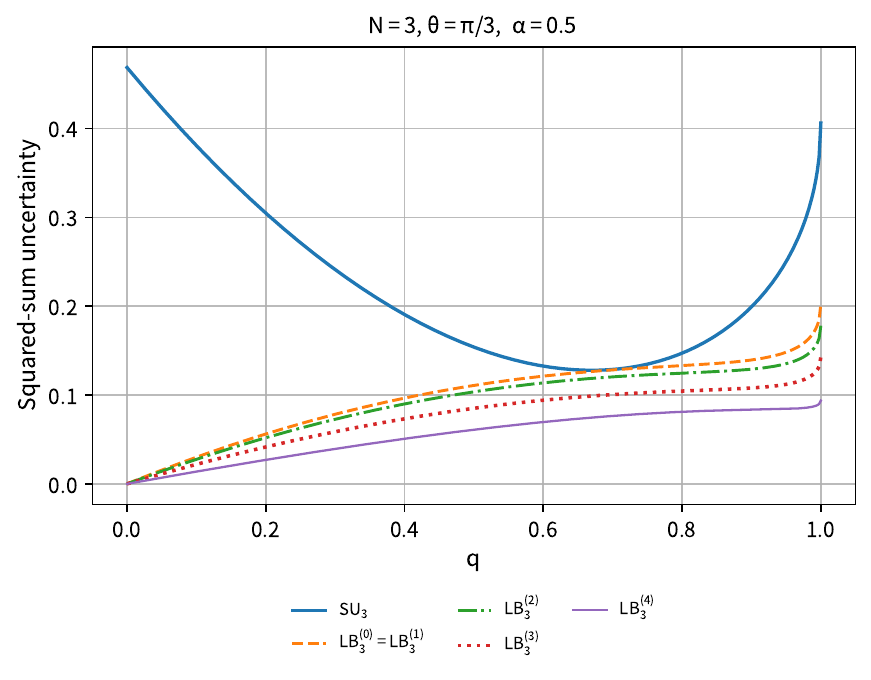}
		\par\smallskip
		\small (c) $N=3$, varying $q$
	\end{minipage}
	\hfill
	\begin{minipage}[b]{0.42\linewidth}
		\centering
		\includegraphics[width=\linewidth]{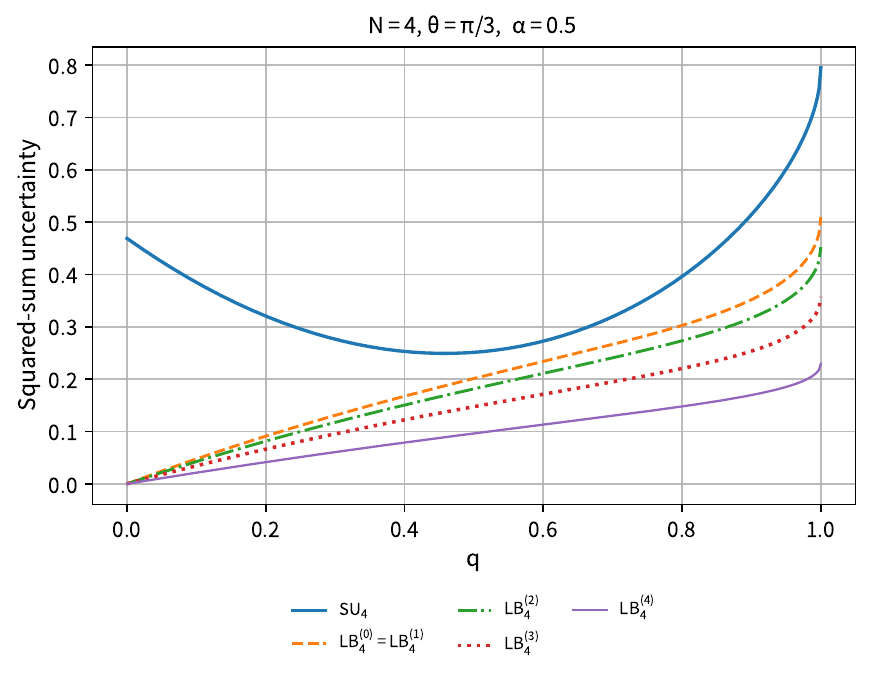}
		\par\smallskip
		\small (d) $N=4$, varying $q$
	\end{minipage}
	
	\vspace{0.3em}
	
	\begin{minipage}[b]{0.42\linewidth}
		\centering
		\includegraphics[width=\linewidth]{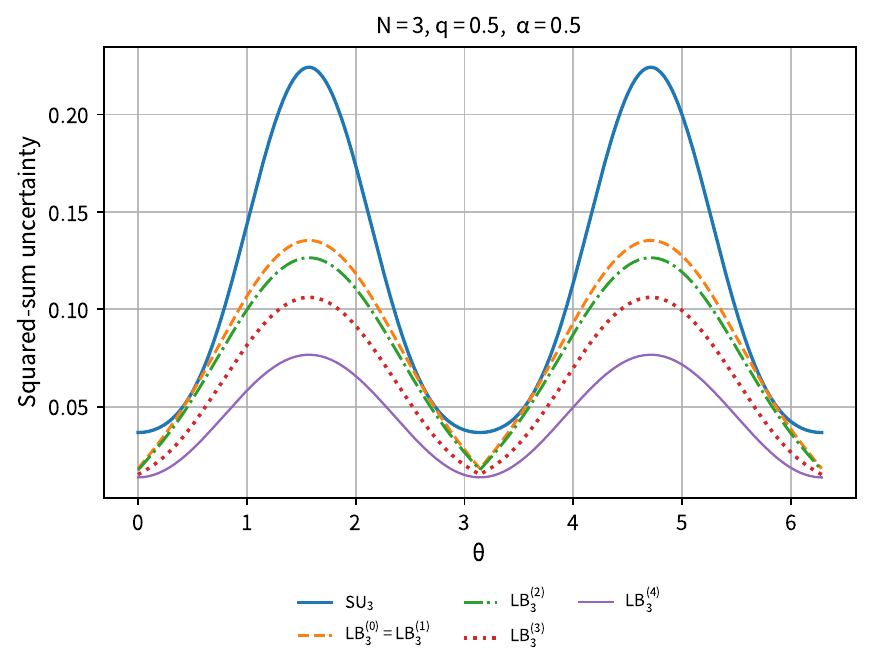}
		\par\smallskip
		\small (e) $N=3$, varying $\theta$
	\end{minipage}
	\hfill
	\begin{minipage}[b]{0.42\linewidth}
		\centering
		\includegraphics[width=\linewidth]{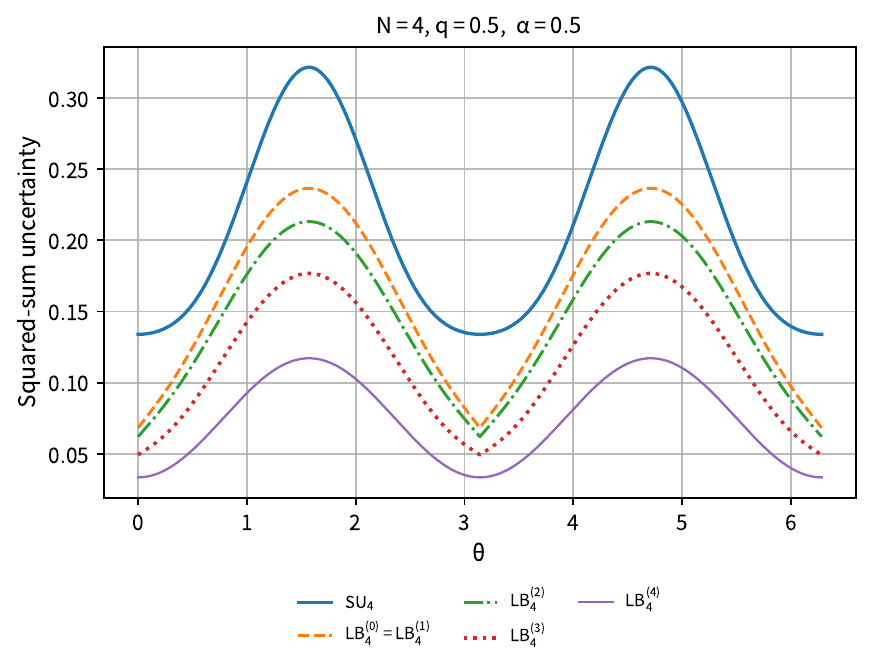}
		\par\smallskip
		\small (f) $N=4$, varying $\theta$
	\end{minipage}
	
\caption{
	Sum of squared uncertainties and the corresponding hierarchy of lower bounds
	for three and four quantum channels.
	The case $N=4$ is obtained by adding the depolarizing channel $\mathcal E_{\mathrm D}(q)$
	to the three channels considered for $N=3$. 
	Panels (a) and (b), (c) and (d), and (e) and (f) show scans in $\alpha$, $q$, and $\theta$, respectively, with $N=3$ on the left and $N=4$ on the right.
	In each scan, the remaining two parameters are fixed
	at their reference values
	$(\theta,q,\alpha)=(\pi/3,1/2,1/2)$.
	The curves represent $\mathrm{SU}_N$
	and the lower bounds $\mathrm{LB}_N^{(k)}$
	for $k=0,1,2,3,4$.
}
\label{fig:N34}
\end{figure}

Define $\SU_N=\sum_sQ_\alpha(\rho,\Phi_s)^2$.
Figure~\ref{fig:N34} shows the dependence on $\alpha$, $q$, and $\theta$, with one parameter varied at a time and the other two held at their reference values $(\theta,q,\alpha)=(\pi/3,1/2,1/2)$.
The parameter ranges are $0.01\leqslant\alpha\leqslant0.99$, $0\leqslant q\leqslant1$, and $0\leqslant\theta\leqslant2\pi$.
For qubit channels, the hierarchy terminates at $p=4$.

Table~\ref{tab:numeric} gives the values at the reference parameters.
For three channels, the $p=2$ bound is $0.103636$, compared with the $p=4$ commutator bound $0.060949$, a relative increase of $70.0\%$ at these parameters.
For four channels, the corresponding values are $0.181480$ and $0.096291$, a relative increase of $88.5\%$.
Both intermediate bounds remain below the $p=0,1$ elementary bound, in agreement with Thm.~\ref{thm:hierarchy}.
At this reference point, the product of the three channel uncertainties is $0.006241$ and the permutation bound in Eq.~\eqref{eq:product} is $0.002612$; the cyclic candidate alone gives $0.001847$.

\begin{table}[tbp]
\caption{Reference point values for $\theta=\pi/3$ and $q=\alpha=1/2$.
All values are computed from Eqs.~\eqref{eq:Q}, \eqref{eq:hierarchy}, and \eqref{eq:comparison} using the stated operator basis and Kraus lists.}
\label{tab:numeric}
\begin{ruledtabular}
\begin{tabular}{lcc}
Quantity & $N=3$ & $N=4$\\
$\SU_N$ & 0.153934 & 0.251156\\
$\LB_N^{(0)}=\LB_N^{(1)}$ & 0.110853 & 0.201304\\
$\LB_N^{(2)}$ & 0.103636 & 0.181480\\
$\LB_N^{(3)}$ & 0.085097 & 0.147414\\
$\LB_N^{(4)}$ & 0.060949 & 0.096291\\
$B_{\Lambda}$ & 0.036146 & 0.058095\\
$B_{\mathrm{hyb}}$ & 0.090719 & 0.116137
\end{tabular}
\end{ruledtabular}
\end{table}

The values of $\Lambda_{st}$ are not invariant under a change of Kraus representation, so Tab.~\ref{tab:numeric} uses only the displayed Kraus lists for this comparison.
The exchange $\alpha\mapsto1-\alpha$ leaves $\rho_\alpha$, and hence every plotted quantity, unchanged.
The hierarchy ordering holds for all parameters by Thm.~\ref{thm:hierarchy}, not merely for the numerical scans.

\section{Conclusion}\label{sec:conclusion}

We have studied product uncertainty relations and bounds on sums of squared uncertainties for quantum channels using the generalized Wigner-Yanase skew information.
The product relation for three channels combines commutator and anticommutator contributions, with an optimization over their possible pairings.
For the sum of squared uncertainties, the inequality for two channels provides the starting point for a general $N$-channel bound involving commutators between Kraus operators of different channels.
The partial Cauchy-Schwarz inequality further gives a hierarchy of lower bounds, allowing the sum of squared uncertainties to be estimated at different levels of refinement.

All levels are independent of Kraus representations, and the endpoint bounds are also independent of the operator basis; intermediate levels can depend on the basis and its ordering.
The first two levels coincide with the elementary arithmetic-geometric mean bound and give no further estimate when the individual channel uncertainties are known.
In the qubit examples considered, the intermediate bounds exceed the commutator bound for the specified states and channel parameters.
The numerical comparisons concern the stated states, channels, and parameter values; those involving the pairing bound also use the specified Kraus lists.

\begin{acknowledgments}
X. H. is supported by the China Scholarship Council and the Research Fund of Jianghan University (Grant No.~2023JCYJ08).
J. Z. acknowledges support from the Doctoral Research Start-up Fund of the University of South China (Grant No.~5524QD125).
N. J. is funded by the Simons Foundation (Grant No.~MP-TSM-00002518).
The authors acknowledge the use of ChatGPT (GPT-6.1, OpenAI) for English language polishing and grammatical corrections.
\end{acknowledgments}


\end{document}